\documentclass{optica-article}

\journal{opticajournal}

\articletype{Research Article}

\usepackage{lineno}

\begin{document}

\title{Space-Time Projection Optical Tomography: Search Space and Orbit Determination}

\author{Hasan Bahcivan\authormark{1} and David Brady\authormark{2} }
\noindent\authormark{1}OpticalX, Tracy, CA, 95304, USA\\
\authormark{2}College of Optical Sciences, University of Arizona, Tucson, AZ, 85751, USA\\

\medskip

\section{Abstract}
In a companion article \cite{bahcivan+etal-2022}, we discussed the radiometric sensitivity
and resolution of a new passive optical sensing technique, Space-Time Projection Optical Tomography (SPOT), to detect and track sub-cm and larger space debris for Space Situational Awareness. SPOT is based on the principle that long synthetic exposure can be achieved if the phase-space trajectory of a hypothetical point-source is precisely predictable within a very wide telescope field-of-view, which is the case for orbiting debris. This article discusses the computational search space for debris mining as well as a recursive measure-and-fit algorithm based on a generalized Hough transform for orbit determination.

\section{Introduction}
 Low Earth Orbit (LEO) already contains millions of debris particles that pose significant collision risk to existing operational spacecraft. Meanwhile, the number of active satellites has been rapidly increasing in the last decade as the commercialization of space from LEO to Geostationary Earth Orbit (GEO) is becoming more attractive for communication and navigation systems.  Considering in-flight breakups and tests of anti-satellite weapons in recent years, debris accumulation will surely accelerate. Since physically de-orbiting such large number of debris is not feasible,  the only means to operate in such a congested space environment is orbit maneuvering and it requires knowing the orbits of each of the  debris pieces as small as mm hours or days ahead of time. 
 
 Tracking of such sub-cm debris is significantly beyond current capabilities. The principal reason for this is the lack of simultaneous scan and stare capability.  Conventional RF and optical systems trade field-of-view (FOV) for sensitivity: increasing aperture improves sensitivity in staring mode but reduces FOV and detection rate. A phased-array, narrow-beam radar such as the Advanced Modular Incoherent Scatter Radar \cite{valentic+etal-2013} can scan fast with only several ms in each beam position and achieve a wide FOV but the sensitivity at each beam position will decrease by the number of beams. Similarly, large-aperture deep-space telescopes are extremely sensitive to detect small debris for population statistics but the rate of detections is very small due to the small FOV.

As a solution that penetrates the cm barrier, we discussed in  \cite{bahcivan+etal-2022} the radiometric sensitivity and resolution of a passive optical technique, Space-time Projection Optical Tomography (SPOT), that computationally overcomes the above scan vs. stare performance trade off. The idea is, instead of conventional single-pixel detections on single frame images, we compute phase-space-pixels (PSP) which are line integrals over hypothetical debris trajectories within the FOV. This method is similar to the generalized Hough transform \cite{hough-1962}. Long synthetic exposures are possible because of the precise analytical description of debris motion along geodesics, i.e., gravity describes free-fall motion for time intervals short enough to ignore unknown thermospheric drag or other non-gravitational forces. PSPs have signal-to-noise ratios (SNR) larger than single frame imaging in proportion to the square root of the number of pixels along the track. Considering an optical array with 10,000-100,000 pixels across, we would expect more than 100X increase in SNR.  Since SNR is proportional to object cross section, such SNR increase via synthetic exposure enables to detect and track objects smaller than 100X in  area or 10X in diameter.

What is remarkable is that, given sufficient computational power, such synthetic exposure can be applied to every phase-space point in the FOV, which means one can stare each and every moving object in the FOV. A massive parallel camera system with a large FOV and massive computational power using  Graphical Processing Units (GPUS)  enables such simultaneous synthetic stare at a large space of phase-space trajectories in 4-6 dimensions.  Most importantly, avoiding the need for flashlight illumination, the Sun illuminates all millions of debris at the same time with its powerful $\sim$1.36 $kW/m^2$ radiation. During twilight times,  a SPOT system can look everywhere within the wide FOV of a survey telescope and can zoom in on each and every particle in the FOV with the sensitivity of a deep-space telescope.

In addition, filtering the data with such extended line integrals (and ultimately with exact TLE projections) presents an advantage to LEO observations with a wide FOV camera by tightly constraining the orbital parameters. Six parameters,  i.e., measurements in $\mathbb{R}^6$  are needed to describe a Keplerian orbit.  While the $\mathbb{R}^4$ assumption for optical observations is largely true for GEO due to the large radial distance to the object and very low angular rates that do not change over the relatively narrow FOV  \cite{milani+etal-2004}, LEO has very high angular rates which change significantly from near zenith to near horizon. Such angular rate variations can be captured by a wide FOV camera design providing the other two undetermined states. Thus optical measurements that enable ${\bf x} \in \mathbb{R}^6$ including angular acceleration significantly constrains the search space enabling the discovery of LEO objects with full orbital parameters.

In this article, we discuss the computational space for various scenarios from straight tracks and intercept-only trajectories to glints as short segments in the FOV. We then briefly discuss orbit determination as a coupled measurement and estimation problem.

\section{The computational search space}

Just as $(i,j)$ uniquely identify a pixel on a rectangular image frame, here we see an orbit as a one unique point in a six-dimensional phase space. If we can properly define the resolution of that one point, we can calculate orbital resolution and the number of grains (resolution cells) we have to sort through to find the object. 

Table \ref{table_grain} (reproduced from \cite{bahcivan+etal-2022}) shows for a test circular orbit the  range of orbital parameters that can be resolved with different track lengths $N$. The resolutions shown are specific to a test circular orbit ($[\theta,\Omega,e, \omega, M, r]=[90^{\circ},0^{\circ}, 0^{\circ}, 0^{\circ}, 0^{\circ}, 15.49962,~{\rm rev/day}]$) and the relative location of the observatory ($[{\rm lat, lon, height}]=[0^{\circ},0^{\circ}, 0~km]$) which was chosen to make the pass an exact overflight at $\approx$ 420 km altitude with the epoch coinciding with the zenith.
Here $[\theta,\Omega,e, \omega, M, r]$ are inclination, longitude of the ascending node, eccentricity, argument of perigee, mean anomaly and revolutions per day, respectively.
To find the resolution for each parameter, the test orbit parameters were incremented, one at a time, until the set of original orbit projection pixels and the new set have 50\% of the pixels common between them. For example, an increment of $\theta$ pivots the projection trajectory around the zenith pixel or incrementing $\Omega$ shifts it laterally. The pivots and shifts would be very different for lower elevation passes or low inclination orbits. 
\begin{table}
\centering
\begin{tabular}{|c c c c c|} 
 \hline
 N (pixels) & 10 & 100 & 1000 & 10000\\ [0.5ex] 
 \hline\hline
 $d\theta$ (deg)& 3.04 & 1.21 & 0.13 & 0.013\\ 
 \hline
 $d\Omega$ (deg) & 0.000180 & 0.000176 & 0.000175 & 0.000174\\
  \hline
 $dr$ (rev/day)& 0.079 &  0.030 & 0.003 & 0.0003\\ 
 \hline
 $dM$ (deg) & 0.000175 &  0.000175 & 0.000175 & 0.000187\\ 
 [1ex] 
 \hline
\end{tabular}
\caption{Computationally-obtained track resolutions for a northbound LEO test circular orbit overflying through the zenith of an optical observatory located at the equator for a range of $NXN$ camera systems with the same IFOV. $dr$ can be converted to altitude resolution $dh$ by multiplying it by $\approx 300$, e.g., $dh\approx 0.1$ km for $N=10000$. 
}
\label{table_grain} 
\end{table}

The complete search space is a 6D mesh with the mesh grains of various sizes which can be precisely determined by the above 50\% rule. Our preliminary calculations (by dividing certain ranges of orbital parameters by the resolutions shown in \ref{table_grain})  show that it could contain more than $10^{20}$  grains per second of observation with $N=$10000. While the exact number can be calculated, searching such large number of tracks does not appear computationally possible at this time \cite{shao+etal-2014, heinze+etal-2019}, hence its practical significance is not clear at this point. Instead, we consider practical search spaces that can be computed entirely as a solution to SSA.

\begin{figure}[h]
\centering\includegraphics[width=10cm]{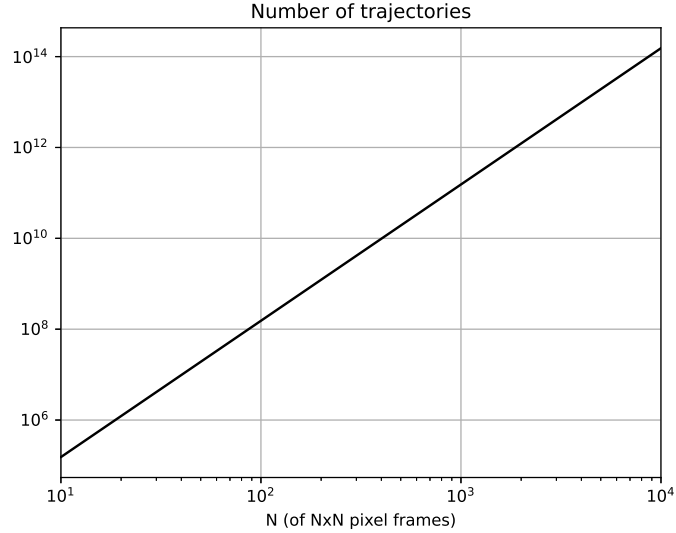}
\caption{Number of all possible distinct straight tracks that can exist at any given moment. Tracks start at a perimeter pixel of an $NXN$ frame and exit at a perimeter pixel of another frame at an angular rate with a lower and upper bound.  The exposure time is set to IFOV divided by the angular rate for a circular orbit at 400 km altitude so that a debris at 400 km transits exactly one pixel per frame). The angular rate is restricted to circular orbits between the altitudes of 350-450 km. Lower (higher) altitude debris transits faster (slower). For example, a 400-km debris entering at the corner of an 100x100 frame and transiting diagonally will exit the FOV at the opposite corner of the 141st frame, while a 450-km debris starting on the same path will exit at the 158th frame.}
 \label{ntrajectories}
\end{figure}

A practical search space is to constrain the search to trajectories short enough to be considered straight. This means that the angular rate along a trajectory is constant and the phase-space-pixel is 4-dimensional. A debris entering the FOV at given a given frame's perimeter pixel and at a given altitude and direction can exit the FOV later only at a perimeter pixel of a particular frame. We therefore count only the entry-exit pixel pairs whose corresponding angular rate falls within the limits of a circular orbit region being searched. This is the same vector-pixel approach used in \cite{heinze+etal-2019}. Figure \ref{ntrajectories} shows the number of such tracks for the altitude range 350-450 km (circular orbit). At a given time, some tracks are initiated at the perimeter and while some tracks reach the perimeter and are terminated. The numbers in Figure \ref{ntrajectories} are that of equilibrium. Note that $\approx 10^{11}$ tracks exist for N=1000 and $\approx 10^{14}$ for N=10000. Considering that Heinze et al. \cite{heinze+etal-2019} reported $1.3 \times 10^{12}$ track computations per hour using the 24-core nodes of the ATLAS computing cluster, N=1000 would be achievable.

The search space could be significantly smaller, however,  if we look only for debris intercepting an operating spacecraft whose orbit is well known. We can use SGP4/SGP4-XP to find hypothetical debris TLE’s that begin within the FOV of the observatory and intercept a particular operational satellite at a future time interval.  No assumption of orbital circularity is needed. 

Consider time now is $t_0$ and an operational satellite will be at position ${\bf r}_{sat}$ at time $t_0+\Delta t$. Consider also that at $t_0$ a debris enters the FOV at the perimeter pixel ($i,j$) at the frame $k_0$, corresponding to the position (${\bf r}_{deb}$). Since we don't know the range to the debris, it will be one of the search variables.  We now invoke the Lambert's problem: determine the interceptor debris orbit ${\bf x}_i$ from the two position vectors (${\bf r}_{sat}(t_0+\Delta t)$, ${\bf r}_{deb}(t_0)$) and the time of flight $\Delta t$: 
\begin{equation}
{\bf x}_i=[\theta,\Omega,e,\omega,M,r]^T = W({\bf r}_{deb},{\bf r}_{sat},\Delta t)
\end{equation}
where $W$ is the Lambert's function.  Then, we apply the following example algorithm for LEO at every frame:
\begin{verbatim}
for dt in range(min_warning_time, time_inc, max_warning_time):
    for each perimeter (i,j) and debris_range in (r_min, dh, r_max):
        r_deb = debris_position(i,j,debris_range)
        r_sat = OrbitPropagator(TLEsat, t0+dt)
        x_i = LambertsFunction(r_deb, r_sat, dt)
        if x_i belongs to LEO:
            [i,j,k] = compute_track_over_fov(x_i)
            SNR = integrate_over([i,j,k])
            if (SNR > SNR_threshold)  
                issue warning
                catalog x_i
\end{verbatim}
Here, the satellite path is partitioned into segments sufficiently small for the intercept to be considered a hazard. Only a small fraction of an orbital period are interceptable, however. For the rest, the debris simply cannot "catch up", unless its orbit is very highly eccentric or parabolic, which could place it out of the LEO category. For example, neglecting the Earth's rotation, a debris departing northbound overhead our optical observatory located on the equator at 0$^{\circ}$ longitude will intercept a similar altitude northbound satellite ascending at 90$^{\circ}$ longitude  at the north pole in 22.5 minutes, assuming 90 min orbital period.  If we restrict the debris altitude within $\pm$50 km of the satellite altitude, its speed must be within 0.37 percent of the satellite's speed. Within this speed range, the particle can intercept the satellite as early as 5 s (from $22.5 min \times \Delta v/v$) before it passes the north pole or as late as 5 s after, an interval of 10 s which is a small fraction ($\approx.2\%$) of the orbital period (90 min).

Now we calculate the number of distinct trajectories that intercept an operational LEO satellite. For each of the perimeter pixels (a total of $4N$) and for each resolvable altitude bin $dh(N)$ (see Table \ref{table_grain}) in an altitude range $H$,  we assume an intercept exists only for $T$ s per orbital period. If $\Delta x$  is the distance below which it is considered an intercept, the number of intercept segments per orbit is $v_{sat}T/ \Delta x$ where $v_{sat}$ is orbital speed. However, we must consider the system resolution. Considering the north pole intercept scenario above, the imager can only resolve limited segments of the target satellite. The length of a resolved segment can be approximated as $\Delta x = v_{sat} dt d\theta (N)$. From Table \ref{table_grain},  for $N=1000$, $d\theta=0.13^{\circ}$. Propagating the orbit for $dt=22.5 min$, we obtain $\Delta x=23$ km. $\Delta x$ will expand further if the intercept is searched for longer warning times. The number of intercept segments is then $(v_{sat}T) / (v_{sat} dt d\theta (N)) = T/(dt d\theta(N))$. Now, the number of intercept-only trajectories $N_{IT}$ per second can be expressed as:
\begin{equation}
N_{IT} = 4N  \times (1/\tau_e) \times (H/dh(N)) \times (T/(dtd\theta(N))
\end{equation}
Note that the factor $1/\tau_e$ is frames per second and there are projectiles launched from the perimeter pixels at every frame.  Assuming the following parameters, $N=1000$, $\tau_e=0.01$s, $H=100$km, $dh=1$ km, $T=10$s, $d\theta=0.13^{\circ}$ and $dt=22.5$ min, we obtain $N_{IT}=1.3e8$/s. This number is to protect only one satellite, however. It must be multiplied by the number of satellites to protect a fleet. For 1000 satellites, the number of tracks will be 10$^11$ about the same order of magnitude as the straight track case above. Note again, no assumptions of orbital circularity is made, the projections are exact.
It is also important to note that once the orbital paths of such satellites are {\it cleared}, they will remain clear for some time until unknown debris fluxes due to orbit perturbations begin diffusing into the protected zones.

Another possibility to narrow the search space is to search for glints. This mode of detection is based on the hypothesis that at various points along the debris trajectory, there will burst of specular reflections off the debris with the number of photons per pixels reaching orders of magnitude larger  than the average per pixel. Such glints would produce short but bright tracklets that can be detected with the aforementioned straight-line integrations but small N. Moreover, because of the likely rotation of the debris and the changing reflection angle, the glints may repeat producing a set of tracklets within the FOV. As discussed below in orbit determination, once one tracklet is detected, it can be linked with the rest in the FOV as part of initial orbit determination. It is also possible to link tracklets from multiple passes using similar tracklet linkage algorithms developed for observing small NEOs with the Space Surveillance Telescope \cite{lue+etal-2019} and in many others. Note the detection methodology used by  \cite{lue+etal-2019} differs from ours, however, as the tracklets were single frame detections whereas ours are phase-space-pixels allowing us an enhanced sensitivity.

Note MEO and GEO will differ from LEO mainly by the time scales and projection trajectory shapes. Debris will spend significantly more time within the FOV. The exposure time will be 100x to 1000x longer.  The frame rate and the data throughput will be less by the same factors. The trajectory shapes, however, far from being straight, will be curved lines or spirals. The exposure rates need to be optimized for these slow motion regions, however. Short exposures used for LEO may amount to unnecessarily high read noise for MEO and GEO. Nevertheless, if some loss of performance can be tolerated, the same collected data can be mined for debris everywhere from LEO to GEO.

\section{Orbit determination}

Here we discuss how we turn measured data into catalog items. We identify three distinct tasks associated with orbit determination (OD): (1) Initial orbit determination (IOD) for uncued debris (2) Precision orbit estimation (POD) for cued debris and (3) Orbit propagation.

IOD  generally refers to the direct computation of six orbital elements from detections during routine data mining for uncatalogued debris. Most IOD methods are based on solving six equations for six orbital elements from two position measurements or three sets of angle measurements. \cite{montenbruck+gill-2000} In SPOT, however, image processing for measurements and orbit determination are inherently coupled and, therefore, treated as a single complete problem. In particular, because the measurements are integrals, {\it a measurement is not complete until the trajectory in the FOV is estimated and the trajectory cannot be estimated fully until the measurement is complete}. The brute-force search approach of integrating over every possible full trajectory is not computationally feasible. Therefore, below we apply a recursive measure and fit method with track extension at every iteration.  Once all iterations are complete, the result corresponds to a single measurement corresponding to a single pass through the FOV. Another note, for the purpose of modeling the FOV projection below, we will describe an orbit with six-parameters only, neglecting non-centric gravitational and non-gravitational forces that must be included for orbit propagation. For the passage time of several minutes over the FOV, the corresponding accelerations for cm and larger objects are too small to cause a deviation in the set of pixels. For sub-cm debris, however, these effects must be properly accounted for, both inside the FOV and outside it for orbit propagation.   

As discussed above, a phase-space-pixel (PSP) is a vector containing a set of parameters from which we can determine, directly or through a coordinate transform, a trajectory that passes through the pixels of a datacube. The PSP intensity value is the sum or weighted sum of those pixel values. A detection is a PSP intensity  exceeding a certain signal-to-noise ratio (SNR) threshold. There is only one detection event per debris per pass, although neighboring PSPs may light up, just like multiple pixels light up around a point source due to the point spread function (PSF) of the optical system. A PSP can correspond to short straight lines, intermediate-length low-order polynomials or long curved lines as direct space-time FOV projection of Earth-centric orbits. The track length depends on how sensitive we want the search to be as the exposure time increases with the track length. The most sensitive search involves whole FOV integrations over exact TLE projections which directly yields TLEs with good precision. However, this is also computationally the most expensive. On the other hand, when short segments return a detection, the associated orbital uncertainty could be very large as seen in Table \ref{table_grain}. Nevertheless, once a short segment is detected, it can be "pulled like a thread" to find the rest of it, with the SNR growing as it is pulled, eventually the length reaching the entire FOV resulting in a precision TLE with a high SNR.

Let ${\bf z}$ denote a straight-line measurement, which is the PSP position of a detection event and assume there are no measurement errors. ${\bf z}$ is equivalent to the measured angular position, e.g., $ { \bf \theta} =[\theta_{az}, \theta_{el}]^T$  of a point source on a conventional image or to the measured radar range $r$ of a point target. Lets say the search using short straight lines returned a detection at ${\bf z}_0=[i_0, j_0,k_0,\alpha_i, \alpha_j, \alpha_k, t_1, t_2]^T$ whose corresponding track pixel positions $(i,j)$ before rounding are given by $[i=i_0 + \alpha_i t, j=j_0 + \alpha_j t, k=k_0 + \alpha_k t]$, where $k$ is the frame number, $(\alpha_i, \alpha_j, \alpha_k)$ are slopes of ($i,j,k$) progressions and $(t_1, t_2)$ are the track start and end times. For example, a stationary source would be represented by ($\alpha_i=0, \alpha_j=0, \alpha_k=1/\tau_e$), whereas streaks on single frames would be represented by $\alpha_{i,j} \gg \alpha_k$. 

We now find in a library $\mathcal{L}$ of coarse-grained orbits a guess orbit ${\bf x}_0=[\theta_0,\Omega_0,e_0,\omega_0, M_0, r_0]^T$ that corresponds to the detected straight line, i.e.,
$$
{\bf z_0}= h[{\bf x}_0],~~~~ {\bf x} \in \mathcal{L}
$$
where $h$ is a transform that projects ${\bf x}$ to ${\bf z}$ in space-time. 
The grain sizes in $\mathcal{L}$ are such that there is one-to-one correspondence between every ${\bf z}$ and ${\bf x}$. For very short tracks, the six-dimensionality of ${\bf x}$ may be reduced to four by setting ($e=0, \omega=0$) corresponding to circular orbits.

The next step is to "pull" the newly found thread by repeating the search with iteratively longer tracks beginning with ${\bf x}_0$. This means we are bringing in more measurements into the estimate, until we reach the end of the track when the object is out of the FOV.
For this, we define the PSP intensity function of track length $L_m$ for the iteration $m$,
$$
Q^{L_m}[h({\bf x})] = \sum_{(i,j,k) \in h({\bf x})}^{L_m} w_{i,j,k}I_{i,j,k} 
$$
which is a weighted sum of the values of the pixels ($i,j,k$) on the projection of ${\bf x}$ on the FOV. The coupled orbit estimation/signal intensification process now attempts to deduce a highly resolved value of ${\bf x}$ that maximizes $Q$. A necessary condition for the intensity function to be maximum with respect to ${\bf x}$ is that 
$$
\frac{ \partial Q^{L_m} } { \partial {\bf x}}=0
$$
A simple solution is to apply the method of gradient descent, also known as steepest descent, although it should be called gradient ascent here. Starting from ${\bf x}_0$,  we take steps proportional to the positive gradient of the function at the current point:
$$
{\bf x}_{n+1} = {\bf x}_n +  \lambda \nabla Q({\bf x}_n)
$$
The gradient descent will quickly approach the solution from a distance, but
its convergence will slow down when it is close to the solution. At some point,  the track length will be increased to $L_{m+1}$ and we repeat the algorithm starting from the last estimate of ${\bf x}$.
This effect of increasing the track length  can be visualized as gradually steepening the $Q$ landscape to prevent the algorithm from "meandering". 
Once $L$ is long enough to span the entire FOV, the final fit ${\bf x}_n^{final}$ becomes the solution. In other words, after utilizing all the data relevant to a single pass, our system is reporting a detection at a particular PSP with the measurement value ${\bf x}_n^{final}$. 

We now calculate the Jacobian for the covariance matrix to determine the uncertainty of the estimated orbit. The above discussion assumes the absence of errors. With errors we have
$$
{\bf x_n^{final}}={\bf x}_{true}+{{\bf \epsilon}(L)}
$$
where ${{\bf \epsilon}(L)}$ represents track-length-dependent measurement errors that depend on a number of factors including the SNR, atmospheric diffraction, or optical aberrations.   Just like we can assign a conventional pixel an angular accuracy or a radar range measurement a range accuracy, we can do the same for a PSP by assigning each of its parameters an uncertainty or we can work with ${\bf \epsilon \epsilon}^T$ to account for correlations between the measurement components. The statistics of measurement errors can be obtained by taking multiple PSP measurements of highly visible spacecraft with precisely-known orbits, e.g., via GPS, then comparing those measurements to the more-precise orbital parameters.  If we assume that ${\bf \epsilon^L}$ is zero-mean Gaussian, we  can form the weighting matrix as ${\bf W}=[E(\epsilon \epsilon^T)]^{-1}$. The orbital uncertainty associated with the final solution ${\bf x}_n^{final}$
can now be computed from the covariance matrix 
$$
C=({\bf J^TWJ})^{-1}
$$
where the Jacobian
$$
J=\frac{\partial Q}{\partial {\bf x} } |_{{\bf x}={\bf x}_n^{final}}
$$
contains the partial derivatives of the modeled observations with respect to the state vector at the reference epoch ${\bf x}_n^{final}$.
The diagonal elements of the covariance matrix yield the standard deviations of the elements of ${\bf x}$ and is a measure of the intervals that most likely contains the actual state.

The second task, POD, is orbit estimation of already-catalogued or partially cued objects to improve a priori orbital elements. The priori information comes in the form of an initial orbit estimate just like ${\bf x}_0$ and the associated covariance matrix that gives a range of orbital parameter values within which a certain high probability of detection exists. Unlike the routine search to detect debris inside the vast 6D space, POD search is an estimation process performed within the much smaller 6D volume of the orbital uncertainties contained in the covariance matrix.

The third task is orbit propagation between observations to provide cues to recapture the object at the next observation point. It is a task beyond the data arc and common to most SSA observing platforms. The most recent orbital prediction model SGP4-XP is available from the government spacetrack.org. SGP4-XP prediction accuracy is 1 to 2 orders of magnitude better than SGP4. SGP4-XP accounts for solar radiation pressure including for high Area-to-Mass-Ratio (HAMR) objects and solar and geomagnetic activity-dependent atmospheric drag. We anticipate that sub-cms will be a totally different regime to account for non-G forces and will require detailed HAMR modeling. 

\section{Conclusion}
Space-Time Projection Optical Tomography is a computational optics solution that enables simultaneous scan and stare capability to search for orbital debris within the wide FOV coverage of a surveillance telescope and with the sensitivity of a deep-space telescope. Compared to single-pixel detections techniques, phase-space-pixel computations dramatically increase the sensitivity of a passive optical system and, to our knowledge, the only means to detect sum-cm debris for effective SSA. While the computational task of searching the 6D space of orbital space appears daunting, certain ranges of sub-spaces exist that are searchable with existing COTS  Graphical Processing Units. We find that   10$^{11}$-10$^{14}$ track computations would provide significant SSA coverage for sub-cm and higher objects for an LEO region with 100 km vertical extent.

\begin{backmatter}
\end{backmatter}


\bibliography{sintraproposal}






\end{document}